\documentclass[10pt,letterpaper]{article}
\usepackage{opex3}
\usepackage{amsmath}
\usepackage{amssymb}
\usepackage{graphicx}
\usepackage[abs]{overpic}
\usepackage{array}
\usepackage{bm}

\newcommand{\bfsfM}{\mbox{\sffamily\bfseries{M}}}
\newcommand{\bfsfT}{\mbox{\sffamily\bfseries{T}}}
\newcommand{\bfsfI}{\mbox{\sffamily\bfseries{I}}}
\newcommand{\bfsfS}{\mbox{\sffamily\bfseries{S}}}
\newcommand{\bfsfa}{\mbox{\sffamily\bfseries{a}}}
\newcommand{\bfsfb}{\mbox{\sffamily\bfseries{b}}}

\begin{document}

%% NOTE: TITLE PAGE & TOC NOT USED FOR MANUSCRIPT SUBMISSIONS %%
%\title{Template and style guide for authors submitting to \textit{Optics Express}}

%\vskip4pc

%\tableofcontents
%\clearpage
%% NO TITLE PAGE FOR OPEX SUBMISSIONS %%

%% START HERE
%%%%%%%%%%%%%%%%%% title page information %%%%%%%%%%%%%%%%%%
\title{Blueshift of the surface plasmon resonance in silver nanoparticles: substrate effects}

\author{S{\o}ren~Raza,$^{1,2}$ Wei~Yan,$^{1,3}$ Nicolas~Stenger,$^{1,3}$ Martijn~Wubs,$^{1,3}$ and N.~Asger~Mortensen$^{1,3,\dag}$}
\address{$^1$Department of Photonics Engineering, Technical University of Denmark, DK-2800 Kgs. Lyngby, Denmark}
\address{$^2$Center for Electron Nanoscopy, Technical University of Denmark, DK-2800 Kgs. Lyngby, Denmark}
\address{$^3$Center for Nanostructured Graphene (CNG), Technical University of Denmark, DK-2800 Kgs. Lyngby, Denmark}

\email{$^\dag$asger@mailaps.org} %% email address is required



\begin{abstract}
We study the blueshift of the surface plasmon (SP) resonance energy of isolated Ag nanoparticles with decreasing particle diameter, which we recently measured using electron energy loss spectroscopy (EELS)~\cite{Raza:2013}. As the particle diameter decreases from 26 down to 3.5~nm, a large blueshift of 0.5~eV of the SP resonance energy is observed. In this paper, we base our theoretical interpretation of our experimental findings on the nonlocal hydrodynamic model, and compare the effect of the substrate on the SP resonance energy to the approach of an effective homogeneous background permittivity. We derive the nonlocal polarizability of a small metal sphere embedded in a homogeneous dielectric environment, leading to the nonlocal generalization of the classical Clausius--Mossotti factor. We also present an exact formalism based on multipole expansions and scattering matrices to determine the optical response of a metal sphere on a dielectric substrate of finite thickness, taking into account retardation and nonlocal effects. We find that the substrate-based calculations show a similar-sized blueshift as calculations based on a sphere in a homogeneous environment, and that they both agree qualitatively with the EELS measurements.
\end{abstract}

 % REPLACE WITH CORRECT OCIS CODES FOR YOUR ARTICLE

%%%%%%%%%%%%%%%%%%%%%%% References %%%%%%%%%%%%%%%%%%%%%%%%%

%\bibliographystyle{osajnl}
%\bibliography{EELS}

%%%%%%%%%%%%%%%%%%%%%%%%%%  body  %%%%%%%%%%%%%%%%%%%%%%%%%%
\section{Introduction}
The use of metal nanoparticles to create astonishing colors in stained glass dates back to ancient Roman times. However, the mechanism behind the color generation was not fully understood until Mie in 1908 rigorously and exactly solved Maxwell's electrodynamical equations for the problem of plane wave scattering off a sphere~\cite{Mie:1908}. From Mie's solution it follows that resonant modes of the metal sphere, which we now refer to as localized SPs~\cite{Maier:2007}, give rise to large absorption cross sections at specific wavelengths, resulting in the colorful stained glass. In Mie's treatment of the problem it is assumed that the material properties of the sphere can be described by a single frequency-dependent function, the local-response dielectric function $\varepsilon(\omega)$. While in most cases a classical treatment based on the dielectric function is justified, important effects due to surface structure~\cite{Lang:1970,Boardman:1975,Bennett:1970,Apell:1981,Feibelman:1982}, nonlocal response~\cite{Ruppin:1973,Boardman:1977,Apell:1982,Schwartz:1982,Fern�ndez-Dom�nguez:2012,David:2012,Toscano:2012,Ciraci:2012} and quantum size effects~\cite{Kreibig:1969,Genzel:1975,Apell:1983,Keller:1993} manifest themselves in the response of metal nanoparticles, when the particle sizes are below $\sim 10$~nm. Many experiments on tiny nanoparticles using both optical measurements~\cite{Kreibig:1985,Charle:1989,Hovel:1993,Tiggesbaumker:1993,Berciaud:2005} and electron energy-loss studies~\cite{Ouyang:1992,Scholl:2012,Raza:2013} have shown that the classical approach is insufficient to describe the experimental observations. The interpretation of these results has been based on semi-classical models, such as the nonlocal hydrodynamic~\cite{Boardman:1982a} and semi-classical infinite barrier (SCIB)~\cite{Apell:1978} approaches, or more complicated quantum calculations using density functional theory~\cite{Lang:1970}.

Recently, we performed EELS on chemically synthesized Ag nanoparticles with diameters ranging from 3.5 to 26~nm~\cite{Raza:2013}. We observed a large blueshift of the localized SP resonance energy from 3.2~eV to 3.7~eV, when the particle size decreased. We interpreted these non-classical observations using two different semi-classical models, the hydrodynamic model and the model presented by Keller et al.~\cite{Keller:1993}, which both only qualitatively could explain the observations. In this paper, we focus on the hydrodynamic model and derive the nonlocal polarizability of a hydrodynamic sphere in a homogeneous environment, which leads to the nonlocal generalization of the Clausius--Mossotti factor. We also study the effect of the substrate on the resonance energy of the nanoparticle. Specifically, we develop an exact formalism to calculate the optical response of a metal sphere on a dielectric substrate of finite thickness, taking into account both retardation and nonlocal response. The theoretical calculations are compared to the EELS measurements.

\section{Experiment: electron energy loss spectroscopy}
The silver nanoparticles are chemically synthesized~\cite{Mulfinger:2007} and afterwards stabilized in an aqueous solution with borohydride ions to prevent aggregation. Subsequently, the solution with nanoparticles is deposited on a plasma-cleaned 10 nm thick Si$_3$N$_4$ TEM membrane purchased from TEMwindows.com. The mean particle diameter is 12 nm with a broad size distribution from 2 nm up to 30 nm, see Fig.~1, which gives us the advantage of being able to perform all of the measurements on the same sample.

The EELS measurements are performed with a FEI Titan transmission electron microscope (TEM) equipped with a monochromator and a probe aberration corrector. The microscope is operated in scanning TEM (STEM) mode at an acceleration voltage of 120~kV, providing a probe diameter of 0.5~nm and a zero-loss peak width of $0.15\pm0.05$~eV. To enhance the excitation of the SP, the EELS spectra are acquired by directing the electron probe to the surface of the silver nanoparticle (aloof trajectory). Details on the data analysis and further experimental information can be found in Ref.~\cite{Raza:2013}.

\begin{figure}
\centering
\begin{overpic}[width=0.9\textwidth]%
{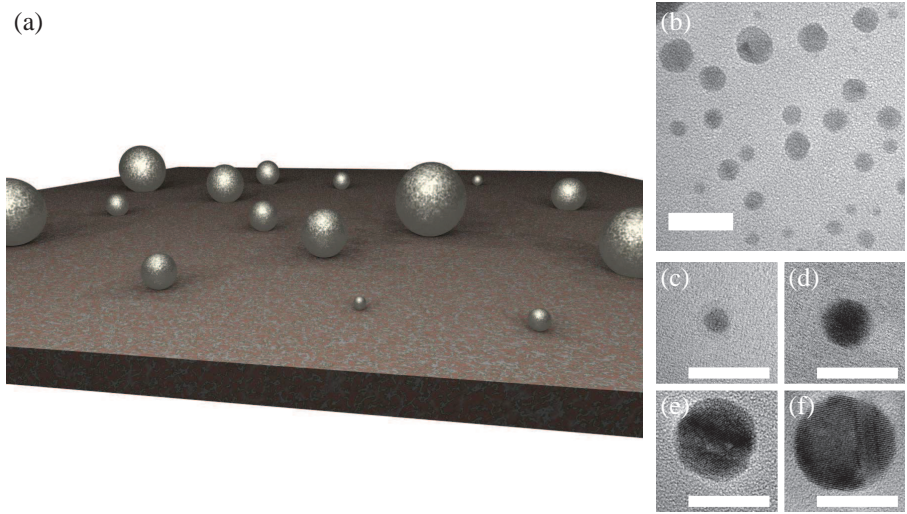}
\put(1,64){(a)}
\put(87,64){{\color{white}(b)}}
\put(87,30){{\color{white}(c)}}
\put(104,30){{\color{white}(d)}}
\put(87,13){{\color{white}(e)}}
\put(104,13){{\color{white}(f)}}
\end{overpic}
\caption{(a) Schematic image of the Ag nanoparticles deposited on 10~nm thick Si$_3$N$_4$ substrate. (b) Bright-field TEM image of sample. (c-f) Bright-field TEM images of single nanoparticles with diameters 3, 6, 10 and 13~nm, respectively. All scale bars are 10~nm long.}
\label{fig:fig1}
\end{figure}

\section{Theory: hydrodynamic model}
In the following theoretical approaches we will assume that the shape of the nanoparticles can be approximated to be spherical. Details and discussion about this approximation can be found in Ref.~\cite{Raza:2013}. Here, we note from the TEM images in Fig.~\ref{fig:fig1}(b-f) that the overall shape of the nanoparticles is spherical, especially for particle sizes below 10~nm in diameter, which justifies our approximation.

We base the interpretation of our experimental results on the hydrodynamic model. We first derive the exact nonlocal polarizability of a metal sphere embedded in a homogeneous material, thereby generalizing the well-known Clausius--Mossotti factor to nonlocal response. The free electrons of the sphere are described by the semiclassical hydrodynamic model, which takes into account nonlocal response but neglects the spill-out of the electrons outside the spheres due to the finiteness of their confining potential. Secondly, the effect of the substrate is taken into account. Here, we present an exact formalism to calculate the retarded optical response of a sphere with hydrodynamic nonlocal response, on a dielectric substrate of finite thickness.

The starting point of the hydrodynamic model is Maxwell's equations in terms of the free-electron density $n$ and free-electron current $\mathbf{J}$~\cite{Boardman:1982a,Boardman:1981,Griffiths:1998}
\begin{subequations} \label{eq:M}
\begin{align}
    \mathbf{\nabla} \cdot \mathbf{D} &= - e n, \label{eq:M1} \\
    \mathbf{\nabla} \cdot \mathbf{H} &= 0, \label{eq:M2} \\
    \mathbf{\nabla} \times \mathbf{E} &= i\omega\mu_0 \mathbf{H}, \label{eq:M3} \\
    \mathbf{\nabla} \times \mathbf{H} &= -i\omega \mathbf{D} + \mathbf{J}, \label{eq:M4}
\end{align}
\end{subequations}
where the constitutive relation $\mathbf{B} = \mu_0 \mathbf{H}$ for non-magnetic materials has been utilized. Here, we introduce the polarization effects due to the bound charges through the constitutive relation for the displacement field $\mathbf{D}=\varepsilon_0 \varepsilon_\infty \mathbf{E}$, where $\varepsilon_\infty$ in general is frequency-dependent and takes into account those polarization effects that are not due to the free electrons, such as interband transitions. The continuity equation, which connects the free-electron density and the free-electron current, follows directly from Eqs.~\eqref{eq:M1} and \eqref{eq:M4},
\begin{align}
    \mathbf{\nabla} \cdot \mathbf{J} = -i\omega e n. \label{eq:CE}
\end{align}
To complete the description of the electromagnetic response of the metal, a relation which connects the free-electron current to the electric field is needed. To this end, we consider the linearized nonlocal hydrodynamic equation~\cite{Boardman:1982a,Bloch:1933a}, which in its real-space formulation becomes~\cite{Toscano:2012,Raza:2011,Toscano:2013a}
\begin{align}
    \frac{\beta^2}{\omega(\omega+i\gamma)}\mathbf{\nabla} \left( \mathbf{\nabla} \cdot \mathbf{J}\right) + \mathbf{J} = \varepsilon_0 \sigma \mathbf{E}, \label{eq:HYD}
\end{align}
where $\sigma = i\omega_\textsc{p}^2/(\omega+i\gamma)$ is the classical Drude conductivity, and $\beta^2=3/5 v_\textsc{f}^2$ with $v_\textsc{f}$ being the Fermi velocity. Within a hydrodynamic description the pressure of the electron gas is included, which gives rise to the presence of compression (longitudinal) waves and leads to spatial dispersion that is observable in truly nanoplasmonic systems. Equations~(\ref{eq:M}-\ref{eq:HYD}) constitute the basic set of equations within the retarded hydrodynamic approach. At an interface between two materials, these equations are supplemented by boundary conditions (BCs). In this study we consider only metal-dielectric interfaces, where Maxwell's BCs must be augmented by a single additional boundary condition (ABC) which states that the normal component of the free-electron current density must vanish~\cite{Boardman:1981,Raza:2011,Sauter:1967,Forstmann:1977,Barton:1979}. The ABC can be derived as a consequence of neglecting the spill-out of electrons.

\subsection{Hydrodynamic sphere in homogeneous environment: nonlocal Clausius--Mossotti factor} \label{sec:ncm}
We consider a small isotropic metal sphere of radius $R$ embedded in a homogeneous dielectric environment with permittivity $\varepsilon_\textsc{b}$. The polarizability $\alpha$ of this sphere is a well-known result in classical optics~\cite{Maier:2007,Jackson:1998} and is given by
\begin{align}
    \alpha=4\pi R^3 \frac{\varepsilon_\textsc{d}-\varepsilon_\textsc{b}}{\varepsilon_\textsc{d}+2\varepsilon_\textsc{b}}, \label{eq:CML}
\end{align}
where $\varepsilon_\textsc{d}=\varepsilon_\infty-\omega_\textsc{p}^2/(\omega^2+i\gamma\omega)$ is the classical Drude permittivity. The factor $(\varepsilon_\textsc{d}-\varepsilon_\textsc{b})/(\varepsilon_\textsc{d}+2\varepsilon_\textsc{b})$ is called the Clausius--Mossotti factor and notice that it is independent of the sphere radius. The polarizability is derived in the quasistatic approximation under the assumption of a static surrounding electric field, thus neglecting spatial variations in the exciting electric field. Our goal is now to derive a generalization to this formula, taking hydrodynamic nonlocal response of the sphere into account. We begin by introducing the electric and current scalar potentials $\phi$ and $\psi$, respectively, defined as
\begin{align}
    \mathbf{E} = - \mathbf{\nabla} \phi, \hspace{1cm} \mathbf{J} = - \mathbf{\nabla} \psi. \label{eq:phipsi}
\end{align}
By inserting Eq.~\eqref{eq:phipsi} into the hydrodynamic Eqs.~(\ref{eq:M}-\ref{eq:HYD}), it can straightforwardly be shown that the scalar potentials inside the metal sphere are governed by the equations~\cite{Villo-Perez:2010}
\begin{subequations} \label{eq:QS}
\begin{align}
    &\left(\nabla^2 + k_\textsc{nl}^2 \right) n = 0, \label{eq:QS1} \\
    &\nabla^2 \phi = \tfrac{e}{\varepsilon_0 \varepsilon_\infty} n, \label{eq:QS2} \\
    &\psi = \tfrac{1}{i\omega-\gamma} \left(\varepsilon_0 \omega_\textsc{p}^2 \phi - e \beta^2 n \right), \label{eq:QS3}
\end{align}
\end{subequations}
where the nonlocal longitudinal wave vector is given as $k_\textsc{nl}^2=(\omega^2+i\omega\gamma-\omega_\textsc{p}^2/\varepsilon_\infty)/\beta^2$. In the surrounding dielectric, the current density $\mathbf{J}$ and electron density $n$ vanish, and the electric scalar potential must instead satisfy the usual Laplace equation $\nabla^2 \phi = 0$. Finally, Maxwell's BCs and the hydrodynamic ABC for the scalar potentials translate into
\begin{align}
    \phi^\text{in} = \phi^\text{out}, \hspace{1cm}
    \varepsilon_\infty \frac{\partial \phi^\text{in}}{\partial r} = \varepsilon_\textsc{b} \frac{\partial \phi^\text{out}}{\partial r}, \hspace{1cm}
    \frac{\partial \psi^\text{in}}{\partial r} = 0, \label{eq:QSBC}
\end{align}
where \textit{in} and \textit{out} refers to inside and outside the metal, respectively. The general solutions to the electric scalar potential and free-electron density inside and outside the sphere are
\begin{subequations} \label{eq:potentials}
\begin{align}
    n^\text{in} &= \sum_{l,m} A_l j_l(k_\textsc{nl}r) Y_{lm}(\theta,\phi), \hspace {1cm} n^\text{out} = 0, \\
    \phi^\text{in} &= \sum_{l,m} \left[D_l r^l - A_l \tfrac{e}{\varepsilon_0 \varepsilon_\infty k_\textsc{nl}^2} j_l(k_\textsc{nl}r) \right] Y_{lm}(\theta,\phi), \\
    \phi^\text{out} &= \sum_{l,m} \left[B_l r^l + C_l r^{-(l+1)} \right] Y_{lm}(\theta,\phi).
\end{align}
\end{subequations}
Here, $j_l$ and $Y_{lm}$ are the spherical Bessel function of the first kind and the spherical harmonics, respectively. The current scalar potential $\psi$ can be determined from Eq.~\eqref{eq:QS3}. We neglect variations in the exciting electric field and assume a constant electric field surrounding the sphere, here directed in the $\mathbf{\hat{z}}$ direction i.e. $\mathbf{E^\text{out}} = E_0 \mathbf{\hat{z}}$. Thus, this poses the requirement that $\lim_{r \to \infty} \phi^\text{out} = -E_0 z = -E_0 r \cos(\theta)$, which excludes all orders of $(l,m)$ in the sums in Eq.~\eqref{eq:potentials} except $(l,m)=(1,0)$. Applying the BCs from Eq.~\eqref{eq:QSBC} and following the usual approach to introducing the polarizability~\cite{Maier:2007}, we determine the nonlocal polarizability $\alpha_\textsc{nl}$ to be
\begin{align}
    \alpha_\textsc{nl}=4\pi R^3 \frac{\varepsilon_\textsc{d}-\varepsilon_\textsc{b}\left(1+\delta_\textsc{nl}\right)}{\varepsilon_\textsc{d}+2\varepsilon_\textsc{b}\left(1+\delta_\textsc{nl}\right)}, \hspace{1cm}
    \delta_\textsc{nl}=\frac{\varepsilon_\textsc{d}-\varepsilon_\infty}{\varepsilon_\infty} \frac{j_1(k_\textsc{nl}R)}{k_\textsc{nl} R j_1'(k_\textsc{nl}R)}, \label{eq:CMNL}
\end{align}
where the prime denotes differentiation with respect to the argument. We see that nonlocal effects enter the Clausius--Mossotti factor as an elegant and simple rescaling of either the metal permittivity from $\varepsilon_\textsc{d}$ to $\tilde{\varepsilon}_\textsc{d}= \varepsilon_\textsc{d} \left(1+\delta_\textsc{nl} \right)^{-1}$ or of the background permittivity from $\varepsilon_\textsc{b}$ to $\tilde{\varepsilon}_\textsc{b} = \varepsilon_\textsc{b} (1+\delta_\textsc{nl})$. Both approaches are equally valid, but we choose to examine the rescaled background permittivity since the nonlocal blueshift of the SP resonance, which is discussed in the following, can be more easily understood in terms of a change in the background permittivity. We point out that the rescaled background permittivity $\tilde{\varepsilon}_\textsc{b}$ is now both frequency- and size-dependent. Finally, we note that when $\beta\rightarrow 0$ then $\delta_\textsc{nl}\rightarrow 0$ in Eq.~\eqref{eq:CMNL} and the classical size-independent Clausius--Mossotti factor is retrieved.

With the nonlocal polarizability we can determine the extinction cross section $\sigma_\text{ext}$ of a metal sphere using the relation~\cite{Maier:2007}
\begin{align}
    \sigma_\text{ext} = \frac{1}{\pi R^2} \left[ \frac{(\omega/c)^4}{6 \pi} |\alpha_\textsc{nl}|^2 + (\omega/c) \text{Im}(\alpha_\textsc{nl}) \right]. \label{eq:ext}
\end{align}
\begin{figure}[!t]
  \centering
  \begin{overpic}[]{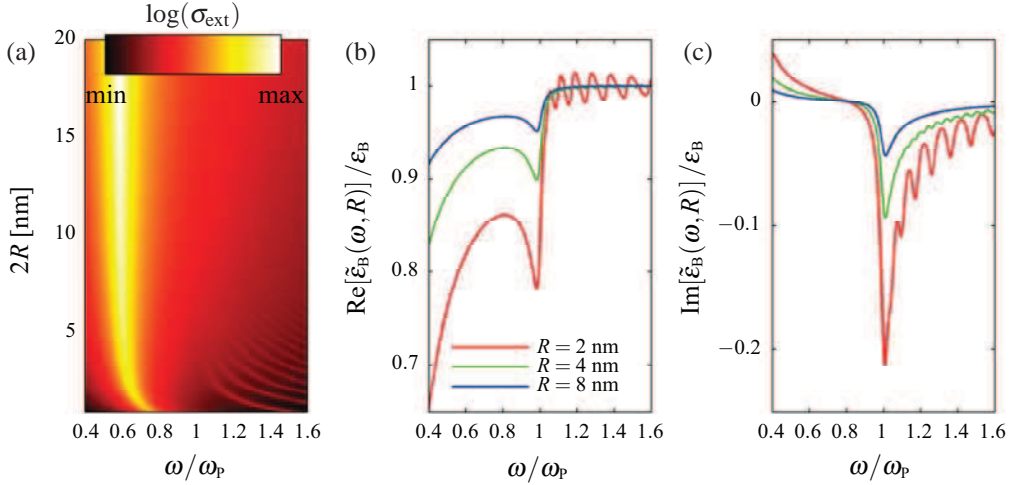}
  \put(19.5,63){$\log(\sigma_\text{ext})$}
  \put(1,58){(a)}
  \put(46,58){(b)}
  \put(91,58){(c)}
  \end{overpic}
  \caption{(a) Extinction cross section based on the nonlocal Clausius--Mossotti factor, Eq.~\eqref{eq:CMNL}, as a function of diameter $2R$ and normalized frequency $\omega/\omega_\textsc{p}$. The real and imaginary parts of the normalized rescaled background permittivity $\tilde{\varepsilon}_\textsc{b}/\varepsilon_\textsc{b}$ as a function of normalized frequency are shown in (b) and (c), respectively, for three different sphere radii: 2~nm (red), 4~nm (green) and 8~nm (blue). Free-electron gas parameters used for the calculations: $\gamma/\omega_\textsc{p}=0.05$, $\beta/c=5\times10^{-3}$, $\varepsilon_\infty = 1$ and $\varepsilon_\textsc{b}=1$.}
  \label{fig:fig2}
\end{figure}

In Fig.~\ref{fig:fig2}(a) we show the extinction cross section as a function of diameter and frequency for a model sphere in vacuum and with only a free-electron response. The blueshift of the SP resonance energy for decreasing particle diameter, which is known to be present from generalized nonlocal Mie theory~\cite{Ruppin:1973}, is captured accurately by the simple nonlocal Clausius--Mossotti factor in Eq.~\eqref{eq:CMNL}. Furthermore, we see that as the particle diameter increases the resonance energy approaches the well-known classical limit $\omega/\omega_\textsc{p} = 1/\sqrt{3} \approx 0.577$. For the smallest diameters ($2R<5$~nm) a series of strongly size-dependent resonances above the plasma frequency can be distinguished. These are resonant pressure-type (longitudinal) waves that arise due to the confinement of the free electron gas. Comparison with the generalized Mie theory~\cite{Ruppin:1973} (not displayed) shows that the spectral location and spectral width of the pressure resonances predicted by the nonlocal Clausius--Mossotti factor are exact.

Using the nonlocal Clausius--Mossotti factor we can deduce a simple approximate, but accurate relation which determines the resonance frequencies of the pressure modes. The poles of the nonlocal correction $\delta_\textsc{nl}$ in Eq.~\eqref{eq:CMNL} determine the spectral position of the pressure modes, which provides us with the condition $j_1'(k_\textsc{nl}R)=0$. We rewrite this condition in terms of the standard Bessel functions and use the large-argument asymptotic form of the Bessel function $J_l(x)\simeq \sqrt{2/(\pi z)} \cos(z-l\pi/2-\pi/4)$, since the product $k_\textsc{nl}R \geq 1$ due to the high frequencies ($\omega>\omega_\textsc{p}$) at which these resonances occur. After some straightforward algebraic manipulations we find (for negligible damping) the relation
\begin{align}
    \omega^2\simeq\frac{\omega_\textsc{p}^2}{\varepsilon_\infty} + \frac{\beta^2\pi^2}{R^2} n^2, \label{eq:bulkres}
\end{align} 
where formally $n=1,2,3,..$. However, upon comparison with extinction cross section calculations we find that the mode $n=1$ is optically dark and therefore does not show up in the extinction spectrum~\cite{Lindau:1970}.

Figures~\ref{fig:fig2}(b-c) display the frequency dependency of the real and imaginary parts of the rescaled background permittivity $\tilde{\varepsilon}_\textsc{b}(\omega, R)$, respectively, for three different radii. In Fig.~\ref{fig:fig2}(b) we see that below the plasma frequency $\text{Re}(\tilde{\varepsilon}_\textsc{b})/\varepsilon_\textsc{b}$ decreases from unity with decreasing radii, leading to the size-dependent blueshift observed in the extinction cross section. In the same frequency interval, we see from Fig.~\ref{fig:fig2}(c) that $\text{Im}(\tilde{\varepsilon}_\textsc{b})/\varepsilon_\textsc{b}$ does not vary significantly and is close to zero. Above the plasma frequency both $\text{Re}(\tilde{\varepsilon}_\textsc{b})/\varepsilon_\textsc{b}$ and $\text{Im}(\tilde{\varepsilon}_\textsc{b})/\varepsilon_\textsc{b}$ display periodic variations, which give rise to the pressure resonances in the extinction cross section. Finally, as the radius increases the frequency dependence of both $\text{Re}(\tilde{\varepsilon}_\textsc{b})/\varepsilon_\textsc{b}$ and $\text{Im}(\tilde{\varepsilon}_\textsc{b})/\varepsilon_\textsc{b}$ weakens, and the classical limits $\text{Re}(\tilde{\varepsilon}_\textsc{b})/\varepsilon_\textsc{b} \rightarrow 1$ and $\text{Im}(\tilde{\varepsilon}_\textsc{b})/\varepsilon_\textsc{b} \rightarrow 0$ are approached.

The above derivation of the nonlocal polarizability $\alpha_\textsc{nl}$ of a metal sphere in a homogeneous dielectric environment is expected to describe many experimental situations of spheres in glass or gels~\cite{Charle:1989,Hovel:1993,Charle:1998}. It can also be used, although its accuracy remains to be tested, in case an inhomogeneous environment is described with an effective homogeneous background dielectric function, see Sec.~\ref{sec:results}.

\subsection{Hydrodynamic sphere on substrate of finite thickness} \label{sec:substrate}
We consider next the case of a metal sphere situated on a substrate, as in the experiment, so we drop the assumption that the background is homogeneous. We present here an exact method based on scattering matrices and multipole expansions to calculate the extinction cross section of the sphere-substrate system, when impinged by a plane wave~\cite{Yan:2013}. The dielectric constant and the thickness of the substrate are denoted $\varepsilon_\textsc{s}$ and $t$, respectively. Taking retardation effects into account, Mie's scattering matrix for the metal sphere is
\begin{equation}
\bfsfT_{l'm'\sigma'}^{lm\sigma}=t_{l}^{\sigma}\delta_{l'm'\sigma'}^{lm\sigma},
\end{equation}
where $\sigma=1,2$ represent TE and TM polarizations, respectively; $\delta_{l'm'\sigma'}^{lm\sigma}=1$ if $l=l'$, $m=m'$, and $\sigma=\sigma'$, otherwise $\delta_{l'm'\sigma'}^{lm\sigma}=0$.
The nonlocal Mie coefficients $t_{l}^{\sigma}$ are given as
\begin{subequations}
\begin{align}
t_{l}^{1}&= -\frac
{{ {j_l}(x_\textsc{d})j_l'(x_\textsc{b})-{j_l}(x_\textsc{b})j_l'(x_\textsc{d})}}
{{{j_l}(x_\textsc{d}) h_l^{(1)\prime}(x_\textsc{b}) - h_l^{(1)}(x_\textsc{b})j_l'(x_\textsc{d})}},\\
t_{l}^{2}&=-\frac
{\left[c_l+j_l'(x_\textsc{d})\right] \varepsilon_\textsc{b} j_l(x_\textsc{b}) - \varepsilon_{\textsc{d}} j_l(x_\textsc{d}) j_l'(x_\textsc{b})}{\left[c_l + j_l'(x_\textsc{d}) \right] \varepsilon_\textsc{b} h_l^{(1)}(x_\textsc{b}) -\varepsilon_{\textsc{d}} j_l(x_\textsc{d}) h_l^{(1)\prime}(x_\textsc{b})},
\end{align}
where $x_\textsc{b}=\omega\sqrt{\varepsilon_\textsc{b}}R/c$ and $x_\textsc{d}=\omega\sqrt{\varepsilon_\textsc{d}}R/c$. The nonlocal correction $c_l$ to the Mie coefficients is given as
\begin{equation}
c_{l}=l(l+1) \frac{j_l(x_\textsc{nl}) j_l(x_\textsc{d})}{ x_\textsc{nl} j_l'(x_\textsc{nl})} \frac{\varepsilon_{\textsc{d}} - \varepsilon_{\infty}}{\varepsilon_{\infty}}, \label{eq:CNL}
\end{equation}
\end{subequations}
with $x_\textsc{nl}=k_\textsc{nl}R$. We note that for $l=1$ the nonlocal correction in Eq.~\eqref{eq:CNL} has the same structural form as $\delta_\textsc{nl}$ in the nonlocal Clausius--Mossotti factor, Eq.~\eqref{eq:CMNL}. In fact they are related as $c_1 = 2 j_1(x_\textsc{d}) \delta_\textsc{nl}$.

In the absence of the substrate, the matrix $\bfsfT$ connects the incident and scattered wave amplitudes $\bfsfa$ and $\bfsfb$, respectively, through the relation $\bfsfb_{l'm'}^{\sigma'}=\bfsfT_{l'm'\sigma'}^{lm\sigma}\bfsfa_{lm}^{\sigma}$. In the presence of the substrate, reflections from the substrate must be taken into account, which changes Mie's scattering matrix $\bfsfT$ to the total scattering matrix $\bfsfM$ given as
\begin{equation}
\bfsfM=(\bfsfI-\bfsfT\bfsfS)^{-1}\bfsfT.
\end{equation}
The total scattering matrix $\bfsfM$ takes into account the interactions between the substrate and the sphere through the substrate scattering matrix $\bfsfS$. The substrate matrix
$\bfsfS$ can be derived by considering the reflections of the spherical waves by the substrate. In particular, $\bfsfS$ is given as
\begin{subequations}
\label{eq:Sexp}
\begin{align}
\bfsfS_{l'm'1}^{lm1}&=s_{ll'}^{mm'}\iint {\text{d}{k_x}\text{d}{k_y}\frac{{f_1y_{lm}^{(1)}y_{l'-m'}^{(1)}+ f_2 }y_{lm}^{(2)}y_{l'-m'}^{(2)}}{{k_{\textsc{b}z} }}},\\
\bfsfS_{l'm'2}^{lm2}&=s_{ll'}^{mm'}\iint {\text{d}{k_x}\text{d}{k_y}\frac{{f_2y_{lm}^{(1)}y_{l'-m'}^{(1)}+ f_1 }y_{lm}^{(2)}y_{l'-m'}^{(2)}}{{k_{\textsc{b}z} }}},\\
\bfsfS_{l'm'1}^{lm2}&=-s_{ll'}^{mm'}\iint {\text{d}{k_x}\text{d}{k_y}\frac{{f_2 }y_{lm}^{(1)}y_{l'-m'}^{(2)}+f_1 y_{lm}^{(2)}y_{l'-m'}^{(1)}}{{k_{\textsc{b}z} }}},\\
\bfsfS_{l'm'2}^{lm1}&=-s_{ll'}^{mm'}\iint {\text{d}{k_x}\text{d}{k_y}\frac{f_1 y_{lm}^{(1)}y_{l'-m'}^{(2)}+f_2 y_{lm}^{(2)}y_{l'-m'}^{(1)}}{{k_{\textsc{b}z} }}},
\end{align}
\end{subequations}
with
\begin{align}
s_{ll'}^{mm'}=\frac{{2{i^{l' - l}}{{( - 1)}^{l + m + m' + 1}}}}{{k_\textsc{b}\sqrt {l'(l' + 1)} \sqrt {l(l + 1)} }}, \quad
y_{lm}^{(1)}=\frac{\partial Y_{lm}(\Omega_{\mathbf k_\textsc{b}})}{\partial \theta_{\mathbf k_\textsc{b}}}, \quad
y_{lm}^{(2)}=\frac{mY_{lm}(\Omega_{\mathbf k_\textsc{b}})}{\sin \theta_{\mathbf k_\textsc{b}}},
\end{align}
where $\mathbf k_\textsc{b}$ represents the wavevector of the plane wave in the background, $k_{\textsc{b}z}$ is the $z$ component of
$\mathbf k_\textsc{b}$ with the imaginary part being non-negative, $\sin \theta_{\mathbf k_\textsc{b}}=\sqrt{k_x^2+k_y^2}/k_\textsc{b}$,
and the integration ranges of $k_x$ and $k_y$ are both from $-\infty$ to $\infty$.
The coefficients $f_{\sigma}$ in Eq. (\ref{eq:Sexp}) represent the reflection coefficients of the substrate for TE and TM
polarized plane waves, respectively, which are expressed as
\begin{subequations}
\begin{align}
f_{\sigma}&=\frac{r_{\sigma}\left[1-\exp(ik_{\textsc{s}z}2t)\right]}{1-r_{\sigma}^2\exp(ik_{\textsc{s}z}2t)}\exp(ik_{\textsc{b}z}2R),
\end{align}
where $k_{\textsc{s}z}$ represents the $z$ component of the wavevector in the substrate. Furthermore, $r_{\sigma}$ is the reflection coefficient between the background and the semi-infinite substrate given as
\begin{equation}
r_1=\frac{k_{\textsc{b}z}-k_{\textsc{s}z}}{k_{\textsc{b}z}+k_{\textsc{s}z}},\quad  r_2=\frac{\varepsilon_\textsc{s}k_{\textsc{b}z}-\varepsilon_\textsc{b}k_{\textsc{s}z}}{\varepsilon_\textsc{s} k_{\textsc{b}z}+\varepsilon_\textsc{b}k_{\textsc{s}z}}.
\end{equation}
\end{subequations}
With the total scattering matrix $\bfsfM$, we can numerically compute the extinction cross section of the metal sphere on a substrate of finite thickness, using the relation
\begin{align}
\sigma_\text{ext}= -\frac{1}{k_\textsc{b}^2} \text{Re}(\mathbf{a}^\text{T} \mathbf{M} \mathbf{a}^*),
\end{align}
where superscripts T and * denote the transpose and complex conjugate, respectively. From the extinction cross section we determine the SP resonance energy.

\begin{figure}[!t]
  \centering
  \begin{overpic}[]{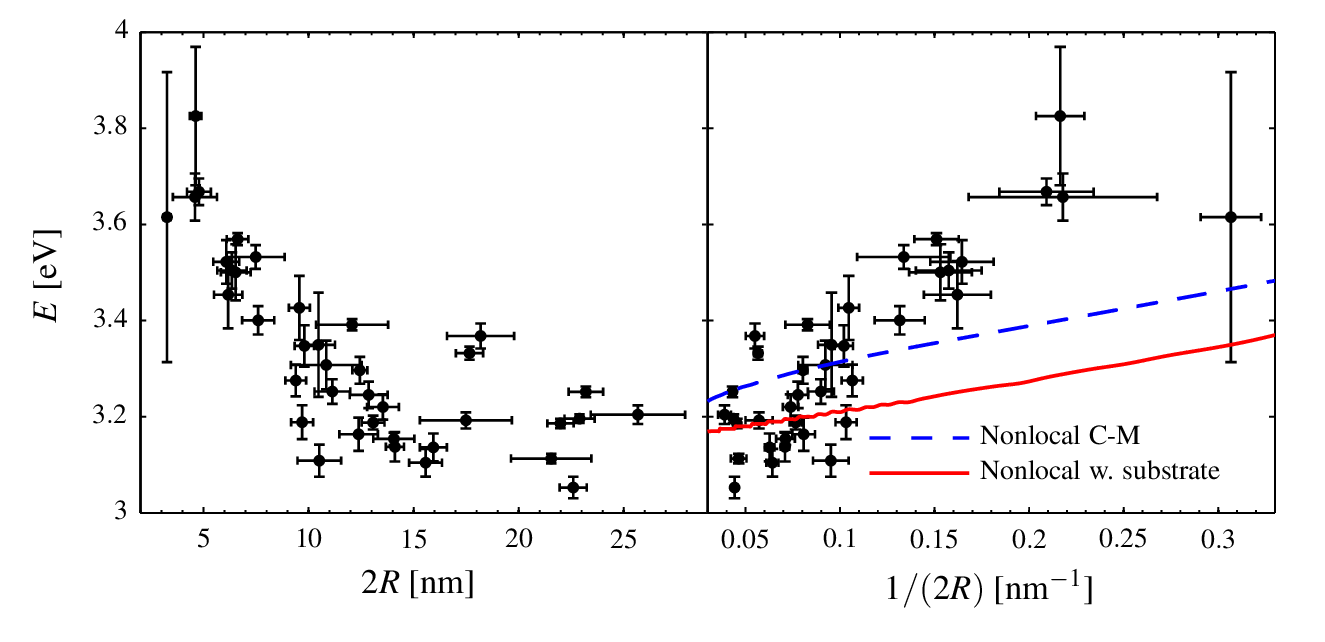}
  \put(22,57){(a)}
  \put(74,57){(b)}
  \end{overpic}
  \caption{EELS measurements of the SP resonance energy $E$ plotted as a function of (a) diameter $2R$ and (b) inverse diameter $1/(2R)$. In (b) the dashed line represents calculations of a nonlocal sphere in a homogeneous environment [nonlocal Clausius--Mossotti factor, Eq.~\eqref{eq:CMNL}]. From the average large-particle ($2R>20~\text{nm}$) resonances we fit $\varepsilon_\textsc{b}=1.53$. The solid line represents calculations of a nonlocal sphere in vacuum situated on a 10~nm thick Si$_3$N$_4$ substrate with permittivity $\varepsilon_\textsc{s}=4.4$~\cite{Baak:1982}. Material parameters for Ag are taken from Ref.~\cite{Rakic:1998a} and the Fermi velocity is $v_\textsc{F}=1.39\times10^6~\text{m/s}$.}
  \label{fig:fig3}
\end{figure}

\section{Results} \label{sec:results}
In Fig.~\ref{fig:fig3}(a), we show the EELS measurements of the SP resonance energy $E$ as a function of the particle diameter $2R$. Two distinct features are present. The first is the spread of the resonance energy at a fixed particle diameter. In Ref.~\cite{Raza:2013} we argue in detail that the spread is due to shape variations of the nanoparticle. Briefly, from the 2D STEM images we determine the area of the particle $A$ and assign it a diameter, assuming a spherical shape $\left(\text{i.e.}~A=\pi R^2\right)$. Different particles with slight deviations from spherical shape can lead to the same area and ultimately the same diameter. However, their SP resonance changes and this is what we observe experimentally. The important second feature we observe is a significant blueshift of the resonance energy of 0.5~eV as the particle diameter decreases. The blueshift is in good agreement with earlier results~\cite{Tiggesbaumker:1993,Ouyang:1992,Scholl:2012}. A classical local-response theory based on a size-independent dielectric function of the material does not predict any frequency shift at all.

Figure~\ref{fig:fig3}(b) displays again the SP resonance energy $E$, now as a function of the inverse particle diameter $1/(2R)$. The experimental measurements suggest a linear relationship between the energy and inverse particle diameter. The nearly linear trend is also seen in the theoretical calculations based on the hydrodynamic model, shown with dashed and solid lines in Fig.~\ref{fig:fig3}(b), albeit with a smaller slope. We point out that the apparent $1/(2R)$ dependent blueshift is only a first-order approximation in the hydrodynamic theory~\cite{Raza:2013}. The dashed line in Fig.~\ref{fig:fig3}(b) corresponds to calculations of a hydrodynamic sphere embedded in a homogeneous environment, i.e. the nonlocal Clausius--Mossotti factor described in Sec.~\ref{sec:ncm}. The permittivity of the background dielectric is fitted to the average resonance of the largest particles ($2R>20~\text{nm}$) to ensure the correct classical SP resonance. We find $\varepsilon_\textsc{b}=1.53$. The solid line shows the resonance energy determined from extinction cross section calculations of a hydrodynamic sphere in vacuum situated on a 10~nm thick Si$_3$N$_4$ substrate, as described in Sec.~\ref{sec:substrate}. Here no fitting of the background permittivity is performed and we use $\varepsilon_\textsc{s}=4.4$ as the permittivity of the substrate, suitable for Si$_3$N$_4$~\cite{Baak:1982}. The same material parameters for the Ag sphere are used in both calculations~\cite{Rakic:1998a}. While the substrate-based calculation shows an overall lower resonance energy for all particle sizes, both approaches show a linear tendency with a nearly identical slope. Compared to a free-space environment the presence of the dielectric substrate should induce a larger blueshift in the hydrodynamic model~\cite{Raza:2013}, and indeed it does (comparison not shown in Fig.~\ref{fig:fig3}). The fitted effective background permittivity in the calculations based on the nonlocal Clausius-Mossotti relation is larger than that of free space, and this makes that the two theoretical curves in Fig.~\ref{fig:fig3}(b) become almost parallel. Especially for the smallest particles $\left[1/(2R)>0.1~\text{nm}^{-1}\right]$ the trend is striking similar, which indicates that (i) only the dipole mode of the sphere is important and (ii) the dipole mode is not significantly altered by the presence of the substrate. However, for larger particle diameters $\left[1/(2R)<0.1~\text{nm}^{-1}\right]$ the substrate alters the dipole mode, which is visible in the slight convex curvature of the solid line in Fig.~\ref{fig:fig3}(b), in contrast to the concave curvature of the dashed line. Surprisingly higher order multipoles in the sphere, which are anticipated to be enhanced due to the presence of the substrate~\cite{Ruppin:1983}, show no significant contribution in the optical response. This is in fact due to the large interband absorption present in Ag at the resonance energies of the higher order multipoles, which heavily dampens the contribution from these modes.

From Fig.~\ref{fig:fig3}(b) we see that the experimentally observed blueshift exceeds the theoretical blueshift predicted by the nonlocal Clausius--Mossotti factor. In Ref.~\cite{Raza:2013} we conjectured that the presence of the substrate could induce the experimentally observed larger blueshift, but from Fig.~\ref{fig:fig3}(b) we see in more detail that the substrate-based calculations do not show a larger shift in energy than the nonlocal Clausius--Mossotti factor. 

\section{Conclusions}
We have studied the experimentally observed blueshift of the SP resonance energy of Ag nanoparticles, when the particle diameters decrease from 26~nm to 3.5~nm. To interpret the measurements we considered two different systems within the theory of the nonlocal hydrodynamic model: a metal sphere embedded in a homogeneous environment and a metal sphere situated on a dielectric substrate of finite thickness. Surprisingly, we find that both systems give rise to similar-sized blueshifts with decreasing particle size, despite the presence of the symmetry-breaking substrate. Both theoretical calculations are in qualitative agreement with the measurements, but the theoretically calculated blueshift is smaller than the blueshift observed in the EELS measurements. Thus, we conclude that the inclusion of the substrate in the theoretical calculations can not quantitatively explain the measurements. This leads us to believe that the deviation between theory and experiment are to be sought for in the intrinsic properties of silver, such as the spill-out of electrons in combination with the screening from the \textit{d} electrons~\cite{Liebsch:1993,Monreal:2013} and size-dependent changes in the electronic band structure~\cite{Hovel:1993}, which are not taken into account in a hydrodynamic description.

\section{Acknowledgements}
We thank Prof. U.~Kreibig and Prof. A.-P.~Jauho for stimulating discussions. We also acknowledge the help and collaboration from our co-authors in Ref.~\cite{Raza:2013}. The Center for Nanostructured Graphene is sponsored by the Danish National Research Foundation, Project DNRF58. The A.~P.~M{\o}ller and Chastine~Mc-Kinney~M{\o}ller Foundation is gratefully acknowledged for the contribution toward the establishment of the Center for Electron Nanoscopy.

\end{document}